%% file: main.tex
\documentclass[ALICE,manyauthors]{cernphprep}
\usepackage[comma,square,numbers,sort&compress]{natbib}
\usepackage{hyperref}
\usepackage{lineno}
\usepackage{xspace}
\usepackage[T1]{fontenc}
\usepackage{orcidlink}

\usepackage{color}

\begin{document}

\input{commands.tex}

\begin{titlepage}
\PHyear{2022}       
\PHnumber{220}      
\PHdate{20 October}  

\title{Search for the Chiral Magnetic Effect with charge-dependent azimuthal correlations in Xe--Xe collisions at $\bf \sqrt{s_{\mathrm{NN}}} = 5.44$~TeV}
\ShortTitle{CME in Xe--Xe collisions}   

\Collaboration{ALICE Collaboration\thanks{See Appendix~\ref{app:collab} for the list of collaboration members}}
\ShortAuthor{ALICE Collaboration} 

\begin{abstract}

Charge-dependent two- and three-particle correlations measured in Xe--Xe collisions at $\sqrt{s_{\mathrm{NN}}} = 5.44$~TeV are presented. Results are obtained for charged particles in the pseudorapidity range $|\eta|<0.8$ and transverse momentum interval $0.2 \le \pt<5.0$~\GeVc{} for different collision centralities. The three-particle correlator $\gc \equiv \gab$, calculated for different combinations of charge sign $\alpha$ and $\beta$, is expected to be sensitive to the presence of the Chiral Magnetic Effect (CME). Its magnitude is similar to the one observed in Pb--Pb collisions in contrast to a smaller CME signal in Xe--Xe collisions than in Pb--Pb collisions predicted by Monte Carlo (MC) calculations including a magnetic field induced by the spectator protons. These observations point to a large non-CME contribution to the correlator. Furthermore, the charge dependence of \gc{} can be described by a blast wave model calculation that incorporates background effects and by the Anomalous Viscous Fluid Dynamics model with values of the CME signal consistent with zero. The Xe--Xe and Pb--Pb results are combined with the expected CME signal dependence on the system size from the MC calculations including a magnetic field to obtain the fraction of CME contribution in \gc{}, $f_{\rm CME}$. The CME fraction is compatible with zero for the 30\% most central events in both systems and then becomes positive. This yields an upper limit of 2\% (3\%) and 25\% (32\%) at 95\% (99.7\%) confidence level for the CME signal contribution to \gc{} in the 0--70\% Xe--Xe and Pb--Pb collisions, respectively.

\end{abstract}
\end{titlepage}

\setcounter{page}{2} 


\section{Introduction} 

The theory of the strong interaction applied to many-body systems predicts that, at sufficiently high densities and temperatures, the protons and neutrons that compose ordinary matter melt into a plasma where quarks and gluons are no longer confined into hadrons. This hot and dense state of matter is called the quark--gluon plasma (QGP)~\cite{Shuryak:1978ij}. The transition from normal hadronic matter to a QGP is supported by Quantum Chromodynamics (QCD) calculations on the lattice~\cite{Karsch:2000kv,Karsch:2000ps,Karsch:2001vs,Allton:2002zi}, where it is found to occur at a temperature of about 155~MeV, and at an energy density $\epsilon$ of about 0.5 GeV/fm$^3$~\cite{Bazavov:2009zn,Bazavov:2011nk,Borsanyi:2010cj}. Collisions between heavy ions accelerated to ultrarelativistic energies can produce the necessary conditions for such a transition to take place~\cite{Chatrchyan:2012mb,Adam:2015lda,Adam:2016thv}.

Heavy-ion collisions may also allow us to access novel QCD phenomena associated with parity violation in strong interactions~\cite{Lee:1973iz,Lee:1974ma,Morley:1983wr,Kharzeev:1998kz,Kharzeev:1999cz,Kharzeev:2015kna,Kharzeev:2007tn,Kharzeev:2007jp,Fukushima:2008xe}. Theoretical expectations indicate that the interaction of quarks with gluonic fields describing transitions between topologically different QCD vacuum states changes the quark chirality and leads to a local chiral imbalance. In the presence of the strong magnetic field produced by the colliding ions~\cite{Skokov:2009qp,Bzdak:2011yy,Deng:2012pc}, this leads to a charge separation (electric current) relative to the reaction plane, the plane defined by the impact parameter and the beam axis. This phenomenon is known as the Chiral Magnetic Effect (CME)~\cite{Fukushima:2008xe}. 

The effects from local parity violation are quantified via the coefficient $a_{1, \alpha}$ in a Fourier decomposition of the particle azimuthal distribution~\cite{Voloshin:1994mz,Voloshin:2004vk}
\begin{equation}
\frac{{\rm d} N}{{\rm d}\Delta\varphi_{\alpha}} \sim 1 + 2v_{1,
  \alpha}\cos(\Delta\varphi_{\alpha}) + 2a_{1,
  \alpha}\sin(\Delta\varphi_{\alpha}) + 2v_{2,
  \alpha}\cos(2\Delta\varphi_{\alpha}) + ... ,
\end{equation}
where $\Delta\varphi_{\alpha} = \varphi_{\alpha} - \Psi_{\rm RP}$, $\varphi_{\alpha}$ is the azimuthal angle of the particle of charge $\alpha$ (+, $-$), and $\Psi_{\rm RP}$ is the reaction plane angle. The coefficients $v_{\rm n, \alpha}$ characterise the anisotropic flow, i.e., the azimuthal anisotropies in particle production relative to $\Psi_{\rm RP}$ due to initial spatial asymmetries of the collision. The degree of overlap between the two colliding nuclei is estimated by the centrality, with low percentage values corresponding to head-on collisions. The first- and second-order flow coefficients ($v_{1,\alpha}$ and $v_{2,\alpha}$) are called directed and elliptic flow, respectively. Since $a_{1, \alpha}$ changes sign from event to event and the average $\langle a_{1,\alpha} \rangle$ over many events is equal to zero, one can only measure $\langle a_{1, \alpha}^2 \rangle$ or $\langle a_{1,+}a_{1,-} \rangle$ that can be accessed through azimuthal correlation techniques. Thus the CME is expected to have an experimentally accessible signal imprinted in the azimuthal correlations between two particles relative to the reaction plane~\cite{Voloshin:2004vk} of the form $\gc \equiv \langle \cos(\varphi_{\alpha} + \varphi_{\beta} - 2\Psi_{\rm RP}) \rangle$. The charge-dependent difference of \gc{} is commonly used to search for the CME. In practice, the reaction plane angle is estimated by constructing the second harmonic symmetry plane angle $\psid$ using azimuthal particle distributions~\cite{Poskanzer:1998yz}, which is why \gc{} is often referred to as a three-particle correlator. The \gc{} correlator measures the difference between the correlations projected onto the reaction plane and perpendicular to it. The contributions from correlations in- and out-of-plane can also be evaluated by measuring the charge-dependent two-particle correlator $\dc \equiv \langle \cos(\varphi_{\alpha} - \varphi_{\beta}) \rangle$.

The first experimental results in Au--Au collisions at a centre-of-mass energy per nucleon--nucleon collision $\sqrt{s_{\mathrm{NN}}} = 200$~GeV at the Relativistic Heavy-Ion Collider (RHIC)~\cite{Abelev:2009ac,Abelev:2009ad} were compatible with initial expectations for the existence of the CME. The subsequent first measurements at the Large Hadron Collider (LHC) in Pb--Pb collisions at $\sqrt{s_{\mathrm{NN}}} = 2.76$~TeV~\cite{Abelev:2012pa} showed a surprising agreement with the results at lower energies, despite the differences in magnitude of the magnetic field~\cite{Skokov:2009qp,Bzdak:2011yy,Deng:2012pc}. Considering that the charged-particle density, \dndeta{}, at the LHC is about three times larger than at RHIC~\cite{Aamodt:2010pb,Aamodt:2010cz}, any signal due to CME will be considerably diluted since it is expected to follow a $1/(\dndeta)$ scaling~\cite{Kharzeev:2007jp}. This effect will be referred to as dilution in the following. The similarity of the two measurements was indicative of the existence of background effects, coming mostly from ``flowing clusters" -- charge-dependent correlations modified by elliptic flow~\cite{Voloshin:2004vk, Wang:2009kd, Bzdak:2009fc, PhysRevC.82.054902}. It was shown in Refs.~\cite{Schlichting:2010qia,Pratt:2010zn} that the local charge conservation coupled to the anisotropic expansion of the medium could explain most if not all the measurements.

To study background effects, the CMS Collaboration performed measurements of charge-dependent correlations in p--Pb collisions at $\sqrt{s_{\mathrm{NN}}} = 5.02$~TeV~\cite{Khachatryan:2016got} and the STAR Collaboration in p--Au and d--Au collisions at $\sqrt{s_{\mathrm{NN}}} = 0.2$~TeV~\cite{STAR:2019xzd}. The results suggest that these correlations are similar to those measured in peripheral Pb--Pb and Au--Au collisions. These results might further indicate the dominance of background effects in peripheral collisions where there is no strong correlation between the magnetic field direction and the orientation of the medium via $\Psi_{\rm RP}$.

These measurements highlighted the need to identify ways of isolating the CME signal from the background. A first attempt was presented by the ALICE Collaboration in Ref.~\cite{Acharya:2017fau} using the Event Shape Engineering method~\cite{Schukraft:2012ah}. This method utilises the fluctuations in the shape of the initial state of the system and allows one to select events with the same centrality but different initial geometry, thus varying the background contributions. The study sets an upper limit of 26--33\% at 95\% confidence level for the CME signal contribution to the charge dependence of \gc{} in the 10--50\% centrality interval. A similar study was performed by the CMS Collaboration~\cite{Sirunyan:2017quh} and the results agree with the measurements in Ref.~\cite{Acharya:2017fau}. A recent study by the ALICE Collaboration~\cite{Acharya:2020rlz} found that charge-dependent correlations relative to the higher harmonic symmetry planes can be used as a proxy for the background, assuming that the correlations relative to $\Psi_2$ and $\Psi_3$ can be factorised. An upper limit of 15--18\% at 95\% confidence level for the CME signal has been reported for the 0--40$\%$ centrality interval, consistent with previous measurements. 

Another approach to address the large backgrounds experimentally is to compare measurements performed in collision systems where the CME contribution is expected to vary significantly, while the background is similar. The STAR Collaboration has recently reported the results of the CME search in an analysis of the three-particle correlator \gc{} measured in collisions of isobar $^{96}_{44}$Ru--$^{96}_{44}$Ru and $^{96}_{40}$Zr--$^{96}_{40}$Zr 
nuclei at $\sqrt{s_{\mathrm{NN}}} = 200$~GeV~\cite{STAR:2021mii}. 
No anticipated CME signature (i.e., a larger magnitude of \gc{} in Ru--Ru than in Zr--Zr collisions due to a larger magnetic field in the former) was observed in that analysis. However, a quantitative analysis taking into account the small geometrical differences between the isobar nuclei is needed for the interpretation of the measurements. One can also try to separate the CME signal and background by comparing the results from  Pb--Pb and Xe--Xe collisions at the LHC since the differences in $v_2$ are typically within 10\% in the 5--70\% centrality interval~\cite{Acharya:2018ihu} but the magnetic field is expected to be significantly larger in Pb--Pb collisions~\cite{Christakoglou:2021nhe}, leading to an increase in the CME contribution. 

In this article, measurements of charge-dependent azimuthal correlations from Xe--Xe collisions at $\sqrt{s_{\rm NN}}$ = 5.44 TeV are presented. The results are compared with earlier measurements in Pb--Pb collisions at $\sqrt{s_{\rm NN}} = 5.02$ TeV~\cite{Acharya:2020rlz} and calculations from a blast wave parameterisation that incorporate background effects and from the Anomalous Viscous Fluid Dynamics (AVFD) model~\cite{Shi:2017cpu,Jiang:2016wve,Shi:2019wzi}. Furthermore, Monte Carlo (MC) simulations of the magnetic field induced by spectator protons with different initial conditions are used to evaluate the expected change in the CME signal between the Xe--Xe and Pb--Pb collisions. This change is then employed to estimate the fraction of the CME signal in both collision systems.

\section{Analysis details}

The data set used for these measurements was recorded with the ALICE detector during the 2017 Xe--Xe run at $\sqrt{s_{\rm NN}}=5.44$ TeV. A detailed overview of the ALICE detector and its performance are available in Refs.~\cite{Aamodt:2008zz, Abelev:2014ffa}. The Inner Tracking System (ITS)~\cite{Aamodt:2010aa}, the Time Projection Chamber (TPC)~\cite{Alme:2010ke}, the V0~\cite{Abbas:2013taa}, and the Zero Degree Calorimeter (ZDC)~\cite{Arnaldi:1999zz}, the main subsystems used in this analysis, are briefly described in the following. The ITS and TPC cover the full azimuth within the pseudorapidity range $|\eta|<0.9$. The ITS consists of six layers of silicon detectors and is employed for tracking, vertex reconstruction, and event selection. The TPC is used to reconstruct charged-particle tracks and to identify particles via specific energy loss, \dEdx{}. The V0 detector, two arrays of 32 scintillator tiles covering $-3.7 < \eta < -1.7$ (V0C) and $2.8 < \eta < 5.1$ (V0A), is used for triggering, event selection, and the determination of centrality~\cite{ALICE-PUBLIC-2018-003} and symmetry plane \psid. Both V0 detectors are segmented in four rings in the radial direction with each ring divided into eight sectors in the azimuthal direction. Two tungsten-quartz neutron ZDCs, installed 112.5 meters from the interaction point on each side, are also used for event selection. 

The trigger conditions and the event selection criteria can be found in Ref.~\cite{Acharya:2018eaq}. Beam-induced background and pileup events are removed using an offline event selection, employing information from the V0, ZDC, and tracking detectors. The primary vertex position is determined from tracks reconstructed in the ITS and TPC as described in Ref.~\cite{Abelev:2014ffa}. Approximately $10^6$ Xe--Xe events in the 0--70\% centrality interval, with a primary vertex position within $\pm 10$~cm from the nominal interaction point along the beam direction, are used in the analysis. The centrality of the collision is estimated from the energy deposition measured in the V0 detector~\cite{ALICE-PUBLIC-2018-003}.

The charged-particle tracks reconstructed using the ITS and TPC within $|\eta|<0.8$ and $0.2 \le \pt < 5.0$~\GeVc{} are used to measure the charge-dependent correlations. Each track is required to have a minimum number of 70 space points (out of a maximum of 159) with a $\chi^2$ per TPC space point lower than 4, to cross at least 70 TPC readout rows, and to have the ratio between the number of crossed rows and the number of findable space points in the TPC larger than 0.8. The selected tracks are also required to have at least 2 ITS hits and a $\chi^2$ per ITS hit smaller than 36. In addition, tracks are selected with a distance of closest approach (DCA) to the reconstructed vertex position smaller than 3.2~cm and 2.4~cm in the longitudinal direction ($z$) and transverse plane ($xy$), respectively. These selection criteria reduce the contamination from secondary charged particles (i.e., particles originating from weak decays, conversions, and secondary hadronic interactions in the detector material) and fake tracks (random associations of space points) and ensure a track momentum resolution better than 4\% in the considered \pt{} interval~\cite{Acharya:2018eaq}. The charged-particle track reconstruction efficiency is estimated from simulations with the HIJING event generator~\cite{Wang:1991hta, Gyulassy:1994ew} combined with the GEANT3 transport model~\cite{Brun:1994aa}. These simulations include a detailed description of the detector response. The \pt{} averaged charge-dependent correlations are corrected for track reconstruction efficiency.

The charge-dependent correlations are measured using two- and three-particle correlators expressed as
\begin{equation}
 \begin{split}
\dc  \equiv \langle \cos(\varphi_{\alpha} - \varphi_{\beta}) \rangle  & =
\langle \cos(\Delta\varphi_{\alpha}) \cos(\Delta\varphi_{\beta})
\rangle + \langle \sin(\Delta\varphi_{\alpha})
\sin(\Delta\varphi_{\beta}) \rangle \\
& = \langle v_{1, \alpha}v_{1, \beta} \rangle + B_{\rm in} + \langle a_{1, \alpha}a_{2, \beta} \rangle + B_{\rm out},
 \end{split}
\end{equation}
\begin{equation}
 \begin{split}
\gc \equiv \langle \cos(\varphi_{\alpha} + \varphi_{\beta} - 2\Psi_{2}) \rangle & = \langle \cos(\Delta\varphi_{\alpha})
\cos(\Delta\varphi_{\beta}) \rangle - \langle
\sin(\Delta\varphi_{\alpha}) \sin(\Delta\varphi_{\beta}) \rangle \\
& = \langle v_{1, \alpha}v_{1, \beta} \rangle + B_{\rm in} - \langle a_{1, \alpha}a_{2, \beta} \rangle - B_{\rm out},
 \end{split}
\end{equation}
where $\Delta\varphi_{\alpha(\beta)} = \varphi_{\alpha(\beta)} - \Psi_{2}$, and $B_{\rm in}$ and $B_{\rm out}$ denote background contributions projected onto \psid{} and perpendicular to it, respectively. The term $\langle v_{1, \alpha}v_{1, \beta} \rangle$ is expected to have negligible charge dependence at midrapidity~\cite{Gursoy:2018yai}. In addition, $\langle v_{1} \rangle$ at midrapidity is zero for a symmetric collision. While \gc{} suppresses background contributions at the level of $v_2$ (i.e., the relative difference between the particle production in-plane and out-of-plane), \dc is dominated by short-range correlations unrelated to \psid{} (``non-flow"), such as inter-jet correlations and resonance decays.

The orientation of the symmetry plane \psid{} is estimated from the azimuthal distribution of the energy deposition measured by the V0A detector, with the $x$ and $y$ components given by
\begin{equation}
\label{eq:Qcomp}  
Q_{\rm 2,x} = \sum_j w_j \cos(2 \varphi_j), \; Q_{\rm 2,y} = \sum_j w_j \sin(2 \varphi_j),
\end{equation}
where the index $j$ runs over the 32 sectors of the V0A detector, $\varphi_j$ is the azimuthal angle of sector $j$ defined by the geometric centre, and $w_j$ is the amplitude of the measured signal in that sector. The symmetry plane resolution is calculated from correlations between the symmetry planes determined with the TPC, the V0A, and the V0C detectors~\cite{Poskanzer:1998yz}. The effect of the decorrelation of $\Psi_2$ between mid and forward pseudorapidity has been estimated to be less than 3\% for $v_2$~\cite{ATLAS:2020sgl}. Any non-uniform detector response is taken into account by adjusting the components of the ${\bf Q}_{2}$ vector using a recentering procedure (i.e.,\ subtraction of the ${\bf Q}_{2}$ vector averaged over many events from the ${\bf Q}_{2}$ vector of each event)~\cite{Selyuzhenkov:2007zi}. The non-flow contributions to the charge-dependent azimuthal correlations are greatly suppressed by the large pseudorapidity separation between the TPC and the V0A ($|\Delta\eta|>2.0$). 

The absolute systematic uncertainties were estimated from the variation of the results with different event and track-selection criteria. The event selection contributions were determined by varying the range of the reconstructed collision vertex position from the nominal interaction point along the beam direction, estimating centrality from the number of hits in the first or second layer of the ITS, and imposing stricter pileup rejection criteria than the default selection. Systematic uncertainties related to track selection criteria were evaluated by changing the ITS hit requirements, varying the minimum number of TPC space points, changing the minimum number of crossed TPC readout rows and the ratio between the number of crossed rows and the number of findable space points in the TPC, rejecting tracks close to the TPC sector boundaries to which the sensitive readout rows do not extend, and comparing any differences between results with only positive and only negative charges for pairs of particles with same charge. Finally, changes of the results due to uncertainties in the tracking efficiency arising from an imperfect description in the simulation of the relative abundances of different particle species and their different reconstruction efficiencies~\cite{ALICE:2018vuu} were considered as part of the systematic uncertainties. The largest contribution to the systematic uncertainties for \gc{} and \dc{} is given by the centrality estimation and track-selection criteria, respectively. The systematic uncertainties are evaluated for each centrality interval. The different sources are assumed uncorrelated and are added in quadrature as an estimate of the total systematic uncertainties if their deviations from the nominal values are significant according to the Barlow criterion~\cite{Barlow:2002yb}. The resulting systematic uncertainties increase from central to peripheral collisions and are summarised in Table~\ref{tab:syst}.
\begin{table}[tp] 
 \centering
 \caption{Summary of absolute systematic uncertainties on the charge-dependent correlations. The uncertainties depend on centrality, whose minimum and maximum values are listed here.}
  \begin{tabular}{ccc}
  \hline
    & Opposite charge & Same charge  \\
  \hline 
  $\dc$  & $(6.8 - 33) \times 10^{-5}$ &  $(3.8 - 13) \times 10^{-5}$ \\  
  $\gc$  & $(1.0 - 8.3) \times 10^{-5}$ &  $(1.4 - 5.9) \times 10^{-5}$ \\   
  \hline
  \end{tabular}
  \label{tab:syst}
\end{table}
%

\section{Results}

\begin{figure}[!tp]
    \begin{center}
    \includegraphics[width =\textwidth]{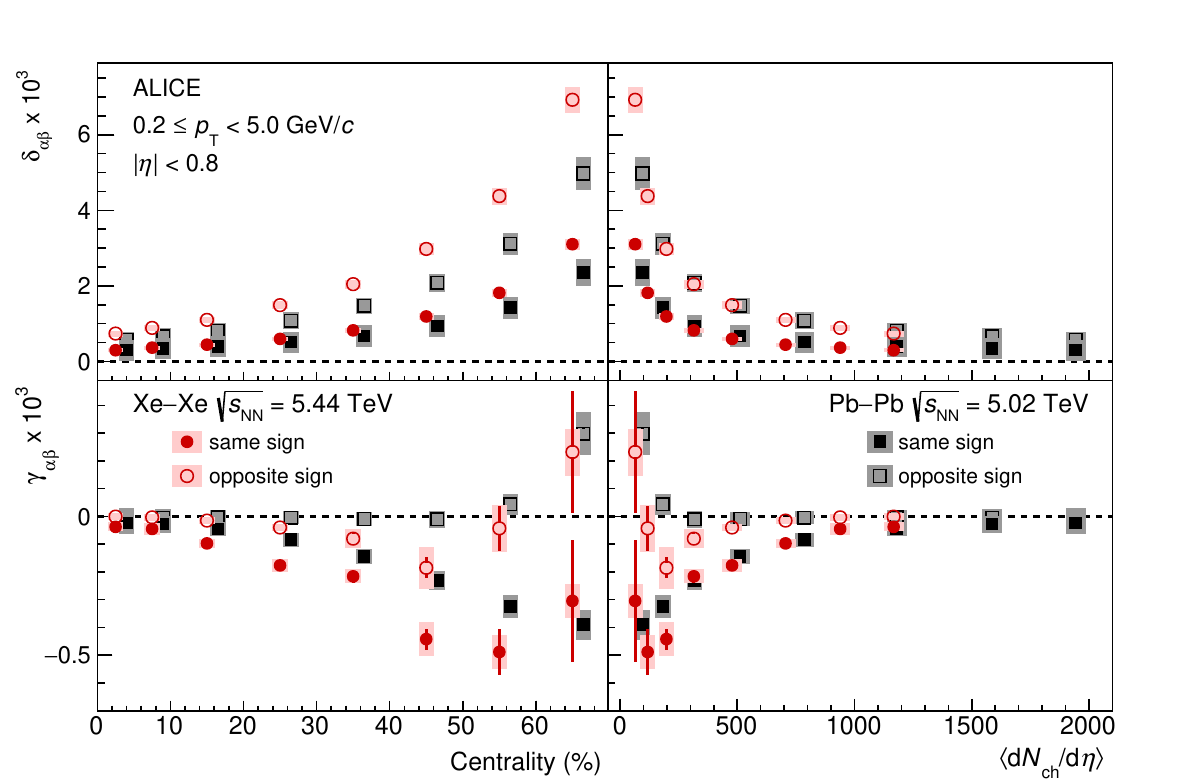}
    \end{center}
    \caption{The \dc{} (top panels) and \gc{} (bottom panels) correlators as a function of centrality (left panels) and charged-particle density~\cite{Adam:2015ptt, Acharya:2018hhy} (right panels) for pairs of particles with same (closed markers) and opposite (open markers) charges from Xe--Xe collisions at $\sqrt{s_{\rm NN}}=5.44$ TeV (red circles) compared to Pb--Pb collisions at $\sqrt{s_{\rm NN}}=5.02$ TeV (black squares)~\cite{Acharya:2020rlz}. The Pb--Pb points are slightly shifted along the horizontal axis for better visibility in the left panels. Bars (boxes) denote statistical (systematic) uncertainties.}
    \label{fig:delta_gamma_compPb}
\end{figure}

Figure~\ref{fig:delta_gamma_compPb} compares the \dc{} and \gc{} correlators for same- and opposite-charge pairs in Xe--Xe collisions at $\sqrt{s_{\rm NN}}=5.44$ TeV to those measured in Pb--Pb collisions at $\sqrt{s_{\rm NN}}=5.02$ TeV~\cite{Acharya:2020rlz} as a function of centrality and average charged-particle multiplicity density \avdndeta{} at midrapidity~\cite{Adam:2015ptt, Acharya:2018hhy}. The results for same-charge pairs denote the average between pairs of particles with only positive and only negative charges since the two combinations are consistent within statistical uncertainties. Both correlators exhibit strong dependence on the charge-sign combination and qualitatively similar centrality dependence in the two systems. For \dc{}, the magnitude of the same- and opposite-charge pair correlations is positive and increases from central to peripheral collisions. In contrast to the CME expectation, the correlation for the opposite-charge pairs is stronger than for the same-charge combinations, indicating that background dominates these measurements. For \gc{}, the magnitude of opposite-charge pair correlations is close to zero within uncertainties for most of the centrality intervals, while it decreases from central to peripheral collisions becoming more negative for same-charge pairs. Thus, the correlation of opposite-charge pairs is weaker than for same-charge pairs. This ordering is compatible with a charge separation with respect to the reaction plane expected in the presence of the CME. 
 
The \dc{} for same-charge pairs shows small (if any) differences between Xe--Xe and Pb--Pb collisions within uncertainties, while the correlations for opposite-charge pairs have larger magnitude in Xe--Xe collisions in the 10--70\% centrality interval. The \gc{} for same- and opposite-charge pairs from Xe--Xe and Pb--Pb collisions have similar magnitudes within uncertainties in the 0--10\% and 50--70\% centrality ranges, while the correlations are stronger in Xe--Xe collisions in the 10--50\% centrality interval. These observations can be attributed to the different number of particles produced in the collision within a given centrality interval between the two systems that dilutes the correlations. This is supported by the fact that the \dc{} and \gc{} for same- and opposite-charge pairs from Xe--Xe collisions are consistent within uncertainties with the corresponding Pb–-Pb results when reported as a function of \avdndeta{}. 
\begin{figure}[tb]
    \begin{center}
    \includegraphics[width=\textwidth]{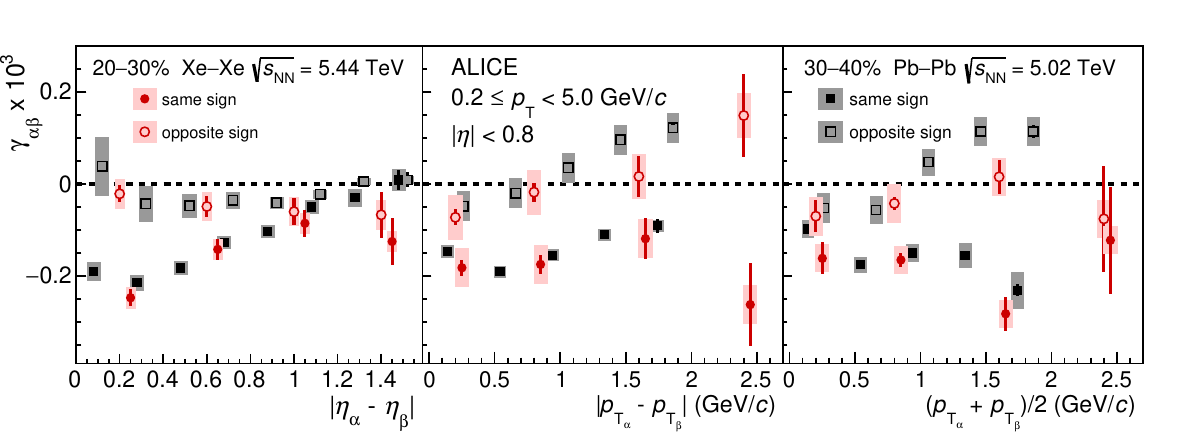}
    \end{center}
    \caption{The dependence of \gc{} on the pseudorapidity difference $|\eta_{\alpha} - \eta_{\beta}|$ (left panel), the transverse momentum difference $|p_{{\rm T}_{\alpha}} - p_{{\rm T}_{\beta}}|$ (middle panel), and the average transverse momentum $(p_{{\rm T}_{\alpha}} + p_{{\rm T}_{\beta}})/2$ (right panel) for pairs of particles with same (closed markers) and opposite (open markers) charges from 20--30\% Xe--Xe collisions at $\sqrt{s_{\rm NN}}=5.44$ TeV (red circles) compared to results from 30--40\% Pb--Pb collisions at $\sqrt{s_{\rm NN}}=5.02$ TeV (black squares)~\cite{Acharya:2020rlz}. The Pb--Pb and Xe--Xe same-charge points are slightly shifted along the horizontal axis for better visibility in all panels. Bars (boxes) denote statistical (systematic) uncertainties.}
    \label{fig:gamma_difAna_2030}
\end{figure}

The \gc{} correlator is also investigated as a function of the pseudorapidity difference $\Delta\eta = | \eta_{\alpha} - \eta_{\beta}|$, the transverse momentum difference $\Delta \pt = |p_{{\rm T}_{\alpha}} - p_{{\rm T}_{\beta}}|$, and the average transverse momentum $\langle \pt \rangle = (p_{{\rm T}_{\alpha}} + p_{{\rm T}_{\beta}})/2$ of the pair. Figure~\ref{fig:gamma_difAna_2030} presents these results for same- and opposite-charge pairs in the 20--30\% centrality interval compared to measurements performed in 30--40\% Pb--Pb collisions~\cite{Acharya:2020rlz}. Different Xe--Xe and Pb--Pb centrality intervals are selected since they have similar transverse densities ($1/S \, \dndeta \sim 10$~fm$^{-2}$ with $S$ being the transverse area) and transverse sizes ($R=\sqrt{S/\pi} \sim 4$ fm)~\cite{Acharya:2018ihu}, thus the contribution from dilution effects is comparable. In addition, the value of $v_2/\varepsilon_{2}$ ($\varepsilon_{2}$ is the second-order eccentricity coefficient and characterises the elliptic shape of the initial geometry) and the influence of radial flow are similar in the two systems for these centrality classes~\cite{Acharya:2018ihu, Acharya:2018eaq}. The opposite-charge pair correlations from Xe--Xe collisions show a weak dependence on $\Delta\eta$ and $\langle \pt \rangle$, while they increase with increasing $\Delta \pt$ of the pair. The correlations for the same-charge pairs do not exhibit any significant dependence on $\Delta \pt$ and $\langle \pt \rangle$ within uncertainties. These correlations show a strong dependence on $\Delta\eta$ with a width of approximately one unit in pseudorapidity difference. The Xe--Xe and Pb--Pb results are compatible within uncertainties demonstrating similar behaviour of this observable despite the differences in the magnetic field.
\begin{figure}[tb]
    \begin{center}
    \includegraphics[width = 0.72\textwidth]{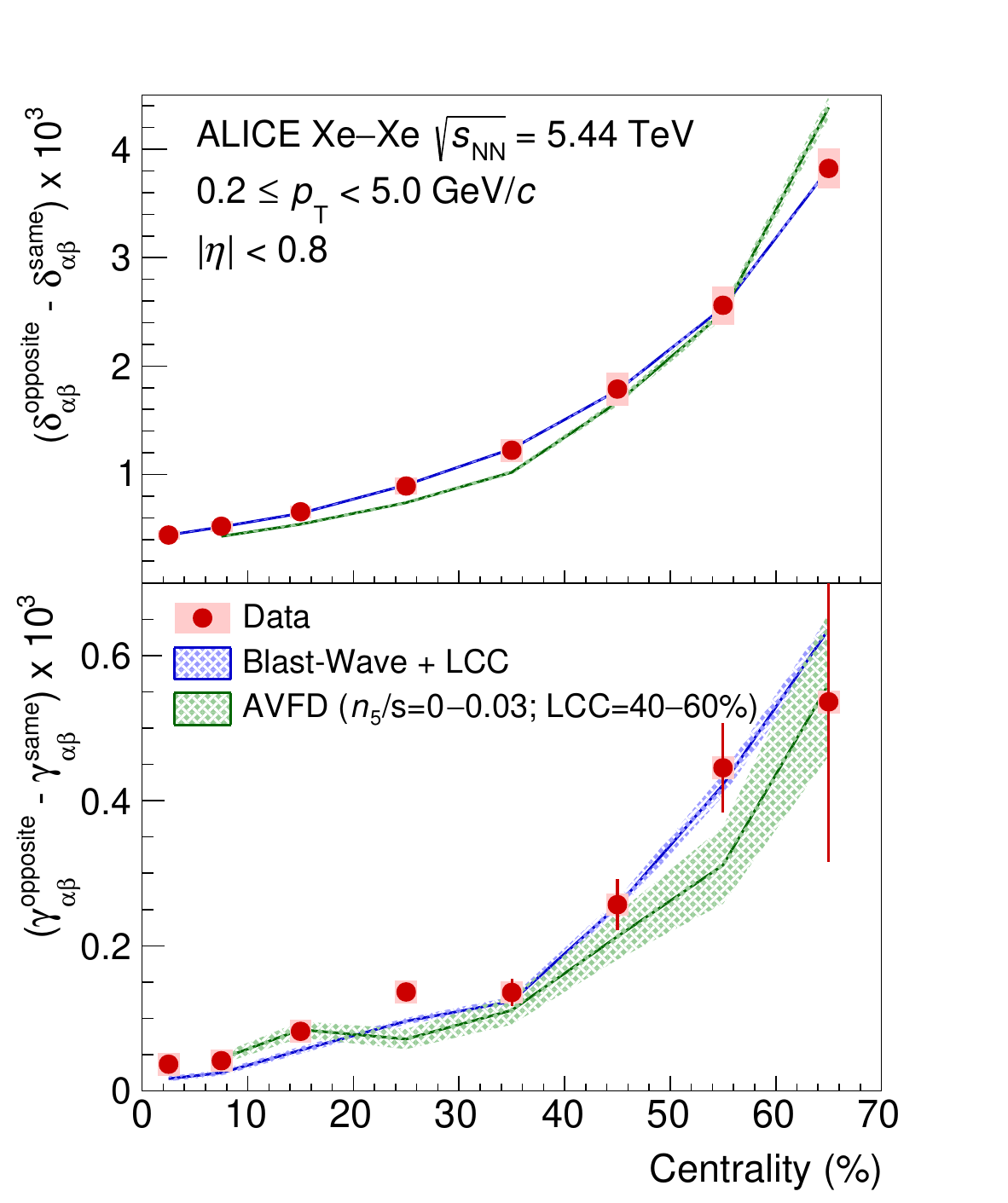}
    \end{center}
    \caption{Centrality evolution of the difference between opposite- and same-charge pair correlations for \dc{} (top panel) and \gc{} (bottom panel) compared to model calculations: blast wave (BW) parameterisation~\cite{Retiere:2003kf} coupled to local charge conservation (LCC) effects (blue curves) and Anomalous Viscous Fluid Dynamics (AVFD)~\cite{Jiang:2016wve,Shi:2017cpu} (green curves). The BW+LCC model is tuned to reproduce the centrality dependence of $\Delta\delta_{\alpha\beta}$, while AVFD is tuned to describe simultaneously the centrality dependence of $\Delta\delta_{\alpha\beta}$ and $\Delta\gamma_{\alpha\beta}$. Bars (boxes) denote statistical (systematic) uncertainties on the data points, while the thickness of the curves represents the uncertainties on the model calculations.}
    \label{fig:DataVsModels}
\end{figure}

To get insight into the origin of the charge-dependent effects observed in Xe--Xe collisions, two different approaches were investigated. The first one relies on a blast wave (BW) model based on a parameterisation from Ref.~\cite{Retiere:2003kf}. Input parameters of the model are tuned to describe the $p_{\rm T}$ spectra~\cite{Acharya:2021ljw} and the $p_{\rm T}$-differential $v_{2}$ values~\cite{ALICE:2021ibz} of charged pions and kaons, as well as of protons and antiprotons, measured in the same collision system and centre-of-mass energy. To account for the main source of background in the measurements reported in this article, the model was further extended by including effects from local charge conservation (LCC). This was done by generating sources uniformly at the surface of the ellipse that surrounds the centre of the system and allowing them to decay into particles with opposite charge. In  this extension of the BW model, the number of sources that emit oppositely-charged pairs is tuned separately for each centrality interval to reproduce the centrality dependence of $\Delta\delta_{\alpha\beta} \equiv \delta^{\mathrm{opp.}}_{\alpha\beta} - \delta^{\mathrm{same}}_{\alpha\beta}$, the correlator that is mainly sensitive to background effects. This is illustrated in 
the upper panel of Fig.~\ref{fig:DataVsModels}, where the BW 
curve is represented by the blue, solid line that goes 
through the experimental data points. The tuned model is then 
used to extract the expectation for the centrality dependence 
of $\Delta\gamma_{\alpha\beta} \equiv \gamma^{\mathrm{opp.}}_{\alpha\beta} - \gamma^{\mathrm{same}}_{\alpha\beta}$, shown in the lower 
panel of Fig.~\ref{fig:DataVsModels}. The width of the band reflects the uncertainty obtained by propagating the corresponding uncertainties of the model parameters using a sub-sampling method. It can be seen that the 
BW model describes fairly well the measured data points for 
all centrality intervals. This is in contrast to the picture 
that emerged in Pb--Pb collisions, where following a similar 
procedure the same model underestimated the measurements 
of $\Delta\gamma_{\alpha\beta}$ by as much 
as $\approx$~40$\%$~\cite{Acharya:2020rlz}.

Additional insight can be obtained by comparing the results 
with calculations from the Anomalous Viscous Fluid Dynamics 
(AVFD) model~\cite{Shi:2017cpu,Jiang:2016wve,Shi:2019wzi}. It relies on a Glauber model description of the initial state of 
the collision and accounts for the development of the early-stage electromagnetic fields as well as for the propagation 
of anomalous fermion currents. The expanding medium is 
described using a 2+1 dimensional viscous hydrodynamics 
(VISH2+1) code~\cite{Song:2007fn} which is coupled to a 
hadron cascade model (UrQMD)~\cite{Bass:1998ca, Bleicher:1999xi}. Within AVFD, the final-state CME signal induced by the initial chirality imbalance is controlled by the axial 
current density $n_5/\mathrm{s}$. At the same time, the 
relevant background in AVFD is governed by the amount of 
positive and negative charged partners emitted from the same 
fluid element relative to the total multiplicity of the event, 
i.e., the LCC percentage. The model was first calibrated 
to describe the centrality dependence of both the 
charged-particle density~\cite{Acharya:2018hhy} and 
the elliptic flow~\cite{Acharya:2018ihu}. As a second step, 
the dependence of $\Delta \delta_{\alpha \beta}$ and 
$\Delta \gamma_{\alpha \beta}$ on both the CME signal and 
the background was determined, by analysing 
samples with either increasing values of $n_5/\mathrm{s}$ or 
LCC percentage, respectively~\cite{Christakoglou:2021nhe}. These results made it possible to extract the combination of $n_5/\mathrm{s}$ and LCC percentage that describes the data and led to a quantitative simultaneous description of the centrality dependence of $\Delta \delta_{\alpha \beta}$ and $\Delta \gamma_{\alpha \beta}$~\cite{Christakoglou:2021nhe}. This is illustrated in the two panels of Fig.~\ref{fig:DataVsModels}. Also in this case, the width of the band reflects the uncertainty obtained by propagating the corresponding uncertainties of the model parameters using a sub-sampling method. The experimental data points can be described by large values of LCC contribution, between 40$\%$ and 60$\%$ for peripheral and more central Xe--Xe collisions, respectively. In addition, the 
values of $n_5/\mathrm{s}$ extracted from this procedure did 
not exhibit any significant centrality dependence and were compatible with zero within uncertainties~\cite{Christakoglou:2021nhe}. Similar to the conclusion extracted from 
the BW model, the study within the AVFD framework indicates 
that the experimental measurements in Xe--Xe collisions 
are dominated by background.
\begin{figure}[tb]
    \begin{center}
    \includegraphics[width = \textwidth]{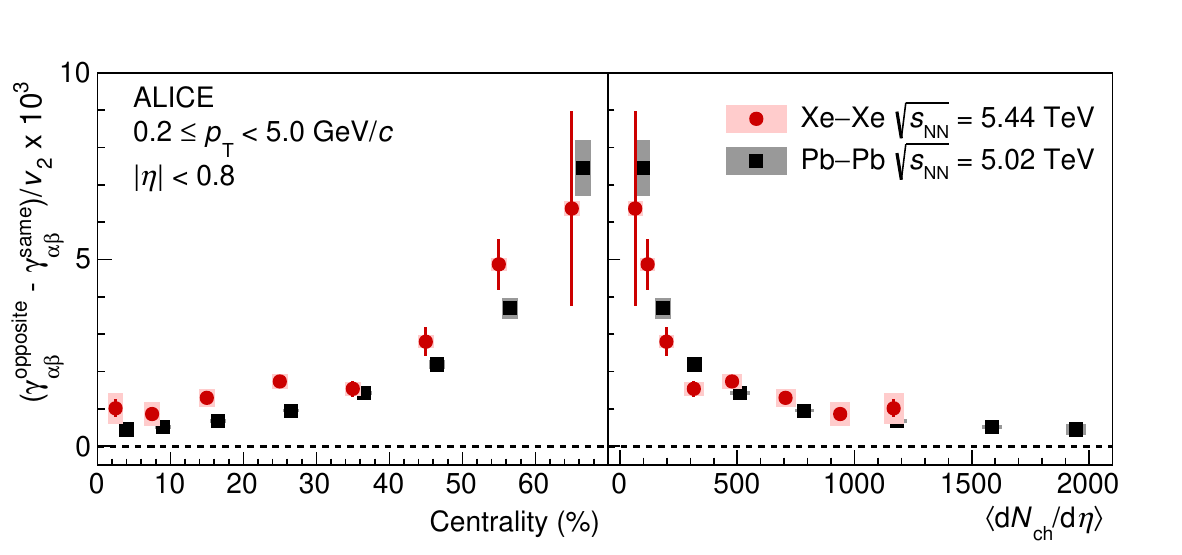}
    \end{center}
    \caption{Difference between opposite- and same-charge pair correlations for \gc{} divided by $v_2$~\cite{Acharya:2018ihu} as a function of centrality (left) and charged-particle density (right) compared to results from Pb--Pb collisions at $\sqrt{s_{\rm NN}} = 5.02$ TeV~\cite{Acharya:2020rlz}. The Pb--Pb points are slightly shifted along the horizontal axis for better visibility in the left panel. Bars (boxes) denote statistical (systematic) uncertainties.}
    \label{fig:gammaDif_compPb}
\end{figure}

The charge separation effect can be further studied using the difference between opposite- and same-charge pair correlations $\Delta\gamma_{\alpha\beta}$ and the CME signal expectations from MC calculations employed as guidance. A comparison between $\Delta\gamma_{\alpha\beta}$ divided by $v_2$~\cite{Acharya:2018ihu} in Xe--Xe collisions and that measured in Pb--Pb collisions~\cite{Acharya:2020rlz} is presented as a function of centrality and charged-particle density~\cite{Adam:2015ptt, Acharya:2018hhy} in Fig.~\ref{fig:gammaDif_compPb}. The value of $\Delta\gamma_{\alpha\beta}/v_2$ is positive for all centralities and its magnitude increases from central to peripheral collisions. Furthermore, it is slightly higher in Xe--Xe than Pb--Pb collisions in the 10--60\% centrality interval. However, the Xe--Xe and Pb--Pb data points fall approximately onto the same curve when reported as a function of \avdndeta{}. 

\begin{figure}[tb]
    \begin{center}
    \includegraphics[width = 0.71\textwidth]{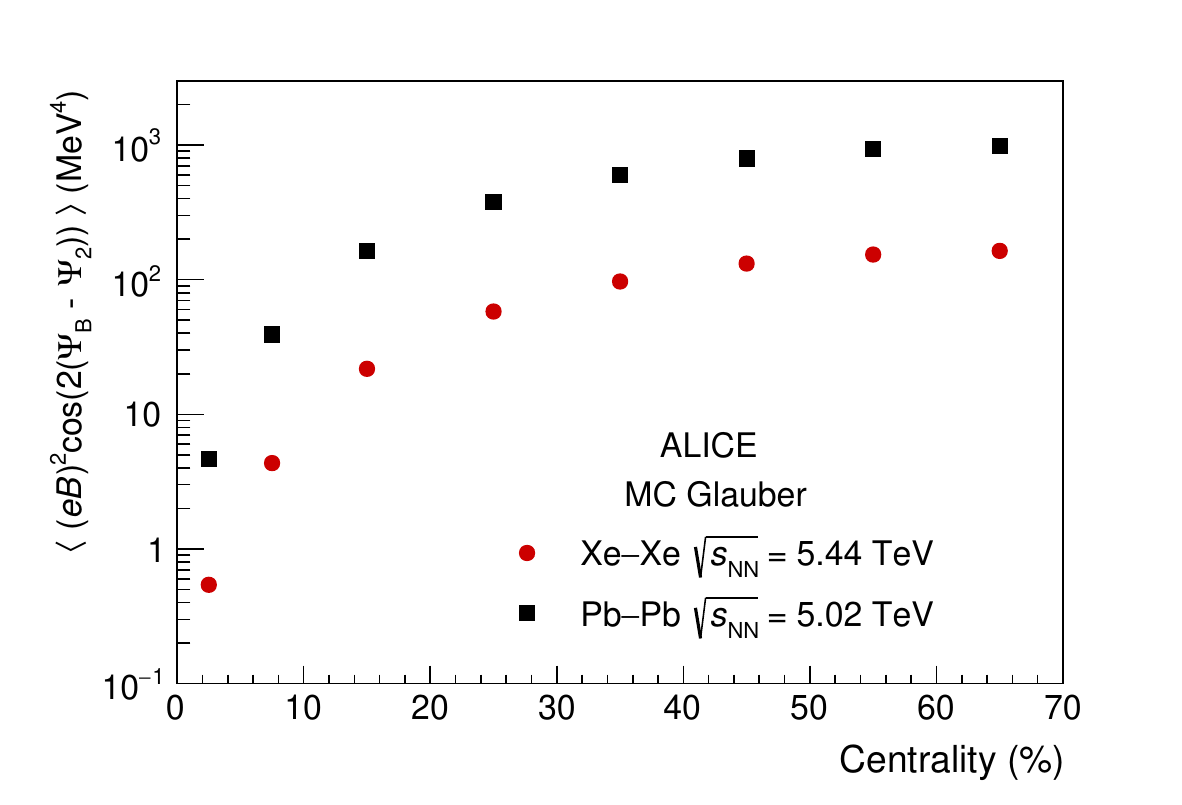}
    \end{center}
    \caption{The expected CME signal as a function of centrality from MC Glauber simulations for Xe--Xe and Pb--Pb collisions~\cite{Miller:2007ri} (see text for details).}
    \label{fig:CME_glb}
\end{figure}

The expected centrality dependence of the CME signal in Xe--Xe and Pb--Pb collisions is estimated from MC Glauber~\cite{Miller:2007ri} simulations including a magnetic field~\cite{Acharya:2017fau}. In these calculations, the $^{208}$Pb nucleus is spherical while the $^{129}$Xe nucleus is deformed with a deformation parameter $\beta_2 = 0.18 \pm 0.02$~\cite{ALICE-PUBLIC-2018-003}. The centrality classes are determined from the multiplicity of charged particles in the acceptance of the V0 detector, which is generated according to a negative binomial distribution with parameters taken from Ref.~\cite{ALICE-PUBLIC-2018-011}. The magnetic field is calculated at the origin from the number of spectator protons using Eq. (A.6) from Ref.~\cite{Kharzeev:2007jp} with the proper time $\tau=0.1$~fm/$c$. Figure~\ref{fig:CME_glb} shows the centrality dependence of $\langle (eB)^2 \cos(2(\Psi_{\rm B} - \Psi_2)) \rangle$ ($\Psi_{\rm B}$ is the direction of the magnetic field $\vec{B}$), which is the expected CME signal contribution in \gc{}, for the two collision systems. The expected CME signal is stronger in Pb--Pb than Xe--Xe collisions in a given centrality interval due to the smaller magnetic field strength and a larger decorrelation between $\Psi_{\rm B}$ and $\Psi_2$ in Xe--Xe collisions. Similar results are found using T$_{\rm R}$ENTo~\cite{Moreland:2014oya} initial conditions. This observation coupled to the agreement of $\Delta\gamma_{\alpha\beta}$ between the two collision systems (see Fig.~\ref{fig:gammaDif_compPb}) points to a large background contribution to \gc{} in Xe--Xe collisions.

The large differences in the CME signal expectations and the small variations in $v_2$ ($<10$\% for the 5--70\% centrality interval)~\cite{Acharya:2018ihu} between the two collision systems can be used to disentangle the potential CME signal from the background in Xe--Xe and Pb--Pb collisions. Assuming that both the CME signal and the background scale with \dndeta{}, the charge dependence of \gc{} for the two collision systems can be expressed using a two-component approach similar to the one proposed in Ref.~\cite{Deng:2016knn}
\begin{align}
    \Gamma^{\rm Xe-Xe} & = s B^{\rm Xe-Xe} + b v_{2}^{\rm Xe-Xe}, \label{eq:g1}\\
    \Gamma^{\rm Pb-Pb} & = s B^{\rm Pb-Pb} + b v_{2}^{\rm Pb-Pb}, \label{eq:g2}
\end{align}
where $\Gamma \equiv \gc \, \dndeta$, $B \equiv \langle (eB)^2 \cos(2(\Psi_{\rm B} - \Psi_2)) \rangle$, and $v_2$ is taken from Ref.~\cite{Acharya:2018ihu} and $\gc^{\rm Pb-Pb}$ from Ref.~\cite{Acharya:2020rlz}. The $s$ and $b$ parameters quantify the signal and background contributions, respectively, and do not depend on collision system within a given centrality interval as a result of the assumption that both scale with \dndeta{}. While the \dndeta{} scaling is expected for $b$ since it is dominated by flowing clusters~\cite{Voloshin:2004vk}, the domains responsible for $s$ can be considered small and thus they can be regarded as ``usual" clusters which scale with \dndeta{}. This scaling is further supported by the AVFD calculations performed for Pb--Pb and Xe--Xe collisions~\cite{Christakoglou:2021nhe}. The $s$ and $b$ parameters can be used to calculate the fractions of the CME signal (denoted as $f_{\rm CME}$) in Xe--Xe and Pb--Pb collisions as
\begin{equation}
    f_{\rm CME} = \frac{sB}{sB + bv_{2}}.
    \label{eq:fcme}
\end{equation}
The smaller CME signal in Xe--Xe collisions also results in a tighter limit on $f_{\rm CME}$ in Xe--Xe than in Pb--Pb collisions. It is worth noting that the CME fractions in the two collision systems are correlated because both are calculated with the same $s$ and $b$ parameters, extracted from the data using Eqs.~\ref{eq:g1} and~\ref{eq:g2}.
\begin{figure}[tb]
    \begin{center}
    \includegraphics[width = \textwidth]{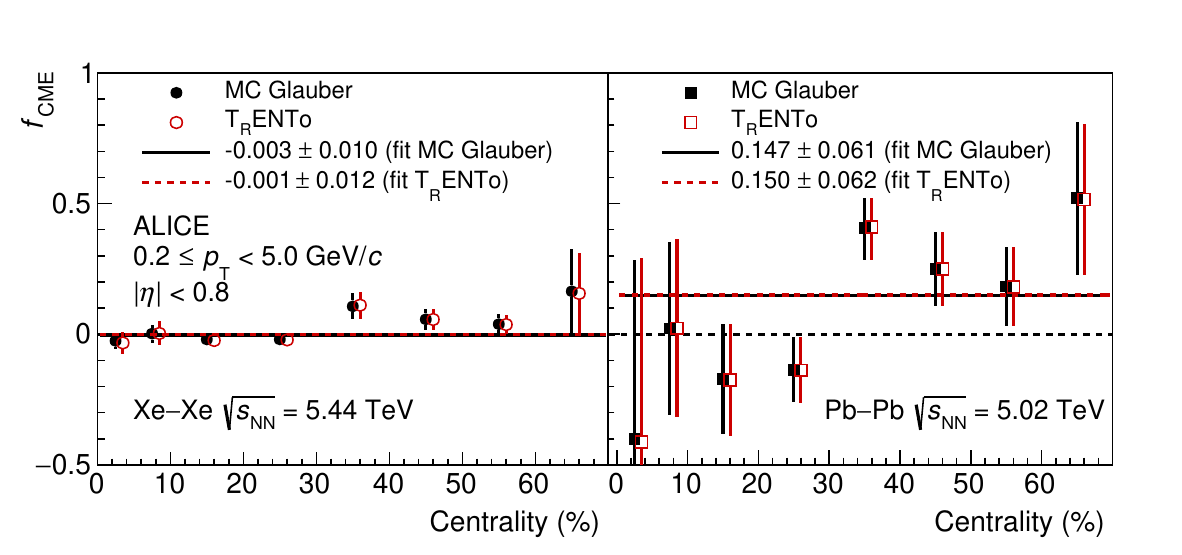}
    \end{center}
    \caption{Centrality dependence of the CME fraction extracted using Eq.~\ref{eq:fcme} with the expected CME signal from MC Glauber~\cite{Miller:2007ri} (closed markers) and T$_{\rm R}$ENTo~\cite{Moreland:2014oya} (open markers) models (see text for details). The T$_{\rm R}$ENTo points are slightly shifted along the horizontal axis for better visibility.}
    \label{fig:fCME_glb_tr}
\end{figure}

Figure~\ref{fig:fCME_glb_tr} presents the centrality dependence of $f_{\rm CME}$ in Xe--Xe and Pb--Pb collisions for the two models used in this study. The uncertainties in the CME fractions are obtained adding in quadrature the statistical and systematic uncertainties in Eqs.~\ref{eq:g1} and~\ref{eq:g2}. The $f_{\rm CME}$ does not depend significantly on the proper time used to calculate the magnetic field since varying the value from 0.1 fm/$c$ to 0.01 fm/$c$, 0.5 fm/$c$, and 1 fm/$c$ yields similar CME fractions. Furthermore, the dependence on the centrality range used in Pb--Pb collisions has been studied. The analysis was performed using only a single centrality interval of Pb--Pb collisions and all Xe--Xe centrality classes. For all eight Pb--Pb centrality intervals, a good agreement is found with the nominal $f_{\rm CME}$. The $f_{\rm CME}$ is compatible with zero up to 30\% centrality in both systems and then becomes positive for midcentral and peripheral collisions with larger values in Pb--Pb than in Xe--Xe. The CME fraction for the 0--30\% centrality interval in Pb--Pb collisions agrees with the one reported in Ref.~\cite{Acharya:2020rlz}. Fitting the data points in the centrality range 0--70\% with a constant function neglecting any centrality dependence gives $f_{\rm CME} = -0.003 \pm 0.010$ ($f_{\rm CME} =-0.001 \pm 0.012$) and $f_{\rm CME} = 0.147 \pm 0.061$ ($f_{\rm CME} = 0.150 \pm 0.062$) for MC Glauber (T$_{\rm R}$ENTo) initial conditions in Xe--Xe and Pb--Pb collisions, respectively. These results are consistent with zero CME fraction in Xe--Xe collisions and correspond to upper limits on $f_{\rm CME}$ of 2\% (3\%) and 25\% (32\%) at 95\% (99.7\%) confidence level for the 0--70\% centrality interval in Xe--Xe and Pb--Pb collisions, respectively. The limits are estimated assuming Gaussian uncertainties.

\section{Summary}

The charge-dependent two- and three-particle correlators \dc{} and \gc{} have been measured in Xe--Xe collisions at $\sqrt{s_{\mathrm{NN}}} = 5.44$~TeV. The charge dependence of these correlators is strongly correlated with centrality, increasing from central to peripheral collisions, and is qualitatively similar to those reported in Pb--Pb collisions. The difference between the Xe--Xe and Pb--Pb results mostly arises from dilution effects since the data points from both collision systems fall approximately onto the same curve when presented as a function of charged-particle density. Monte Carlo simulations with different initial state models predict a significantly larger magnitude of the CME signal in Pb--Pb than Xe--Xe collisions, which implies that the dominant contribution to \gc{} in Xe--Xe collisions is due to background effects. The magnitude of the charge dependence of \gc{} is described over the entire centrality range by a blast wave parameterisation that incorporates local charge conservation tuned to reproduce the components of the background. This magnitude is also reproduced by Anomalous Viscous Fluid Dynamics calculations with large contributions from local charge conservation effects and values of the CME signal close to zero, thus indicating that the background is the dominant contribution to the three-particle correlator. In order to get a quantitative estimate of the signal and background contributions, the measured values of \gc{} in Xe--Xe and Pb--Pb collisions are compared using a two-component approach. This procedure allows one to estimate the fraction of the CME signal in both collision systems. Averaging over the 0--70\% centrality interval, an upper limit of 2\% (3\%) and 25\% (32\%) is estimated at 95\% (99.7\%) confidence level for the CME contribution to the charge dependence of \gc{} in Xe--Xe and Pb--Pb collisions, respectively.


\newenvironment{acknowledgement}{\relax}{\relax}
\begin{acknowledgement}
\section*{Acknowledgements}
\input{fa_2022-10-11_Opt_C.tex}
\end{acknowledgement}

\bibliographystyle{utphys}   
\bibliography{bibliography}

\newpage
\appendix

%
%

\section{The ALICE Collaboration}
\label{app:collab}
\input{2022-10-11-Alice_Authorlist_2022-10-11_Opt_C.tex}
\end{document}

%% file: commands.tex
%

\newcommand{\pp}           {pp\xspace}
\newcommand{\ppbar}        {\mbox{$\mathrm {p\overline{p}}$}\xspace}
\newcommand{\XeXe}         {\mbox{Xe--Xe}\xspace}
\newcommand{\PbPb}         {\mbox{Pb--Pb}\xspace}
\newcommand{\pA}           {\mbox{pA}\xspace}
\newcommand{\pPb}          {\mbox{p--Pb}\xspace}
\newcommand{\AuAu}         {\mbox{Au--Au}\xspace}
\newcommand{\dAu}          {\mbox{d--Au}\xspace}

\newcommand{\s}            {\ensuremath{\sqrt{s}}\xspace}
\newcommand{\snn}          {\ensuremath{\sqrt{s_{\mathrm{NN}}}}\xspace}
\newcommand{\pt}           {\ensuremath{p_{\rm T}}\xspace}
\newcommand{\meanpt}       {$\langle p_{\mathrm{T}}\rangle$\xspace}
\newcommand{\ycms}         {\ensuremath{y_{\rm CMS}}\xspace}
\newcommand{\ylab}         {\ensuremath{y_{\rm lab}}\xspace}
\newcommand{\etarange}[1]  {\mbox{$\left | \eta \right |~<~#1$}}
\newcommand{\yrange}[1]    {\mbox{$\left | y \right |~<~#1$}}
\newcommand{\dndy}         {\ensuremath{\mathrm{d}N_\mathrm{ch}/\mathrm{d}y}\xspace}
\newcommand{\dndeta}       {\ensuremath{\mathrm{d}N_\mathrm{ch}/\mathrm{d}\eta}\xspace}
\newcommand{\avdndeta}     {\ensuremath{\langle\dndeta\rangle}\xspace}
\newcommand{\dNdy}         {\ensuremath{\mathrm{d}N_\mathrm{ch}/\mathrm{d}y}\xspace}
\newcommand{\Npart}        {\ensuremath{N_\mathrm{part}}\xspace}
\newcommand{\Ncoll}        {\ensuremath{N_\mathrm{coll}}\xspace}
\newcommand{\dEdx}         {\ensuremath{\textrm{d}E/\textrm{d}x}\xspace}
\newcommand{\RpPb}         {\ensuremath{R_{\rm pPb}}\xspace}

\newcommand{\nineH}        {$\sqrt{s}~=~0.9$~Te\kern-.1emV\xspace}
\newcommand{\seven}        {$\sqrt{s}~=~7$~Te\kern-.1emV\xspace}
\newcommand{\twoH}         {$\sqrt{s}~=~0.2$~Te\kern-.1emV\xspace}
\newcommand{\twosevensix}  {$\sqrt{s}~=~2.76$~Te\kern-.1emV\xspace}
\newcommand{\five}         {$\sqrt{s}~=~5.02$~Te\kern-.1emV\xspace}
\newcommand{\twosevensixnn}{$\sqrt{s_{\mathrm{NN}}}~=~2.76$~Te\kern-.1emV\xspace}
\newcommand{\fivenn}       {$\sqrt{s_{\mathrm{NN}}}~=~5.02$~Te\kern-.1emV\xspace}
\newcommand{\LT}           {L{\'e}vy-Tsallis\xspace}
\newcommand{\GeVc}         {Ge\kern-.1emV/$c$\xspace}
\newcommand{\MeVc}         {Me\kern-.1emV/$c$\xspace}
\newcommand{\TeV}          {Te\kern-.1emV\xspace}
\newcommand{\GeV}          {Ge\kern-.1emV\xspace}
\newcommand{\MeV}          {Me\kern-.1emV\xspace}
\newcommand{\GeVmass}      {Ge\kern-.2emV/$c^2$\xspace}
\newcommand{\MeVmass}      {Me\kern-.2emV/$c^2$\xspace}
\newcommand{\lumi}         {\ensuremath{\mathcal{L}}\xspace}

\newcommand{\ITS}          {\rm{ITS}\xspace}
\newcommand{\TOF}          {\rm{TOF}\xspace}
\newcommand{\ZDC}          {\rm{ZDC}\xspace}
\newcommand{\ZDCs}         {\rm{ZDCs}\xspace}
\newcommand{\ZNA}          {\rm{ZNA}\xspace}
\newcommand{\ZNC}          {\rm{ZNC}\xspace}
\newcommand{\SPD}          {\rm{SPD}\xspace}
\newcommand{\SDD}          {\rm{SDD}\xspace}
\newcommand{\SSD}          {\rm{SSD}\xspace}
\newcommand{\TPC}          {\rm{TPC}\xspace}
\newcommand{\TRD}          {\rm{TRD}\xspace}
\newcommand{\VZERO}        {\rm{V0}\xspace}
\newcommand{\VZEROA}       {\rm{V0A}\xspace}
\newcommand{\VZEROC}       {\rm{V0C}\xspace}
\newcommand{\Vdecay} 	   {\ensuremath{V^{0}}\xspace}

\newcommand{\ee}           {\ensuremath{e^{+}e^{-}}} 
\newcommand{\pip}          {\ensuremath{\pi^{+}}\xspace}
\newcommand{\pim}          {\ensuremath{\pi^{-}}\xspace}
\newcommand{\kap}          {\ensuremath{\rm{K}^{+}}\xspace}
\newcommand{\kam}          {\ensuremath{\rm{K}^{-}}\xspace}
\newcommand{\pbar}         {\ensuremath{\rm\overline{p}}\xspace}
\newcommand{\kzero}        {\ensuremath{{\rm K}^{0}_{\rm{S}}}\xspace}
\newcommand{\lmb}          {\ensuremath{\Lambda}\xspace}
\newcommand{\almb}         {\ensuremath{\overline{\Lambda}}\xspace}
\newcommand{\Om}           {\ensuremath{\Omega^-}\xspace}
\newcommand{\Mo}           {\ensuremath{\overline{\Omega}^+}\xspace}
\newcommand{\X}            {\ensuremath{\Xi^-}\xspace}
\newcommand{\Ix}           {\ensuremath{\overline{\Xi}^+}\xspace}
\newcommand{\Xis}          {\ensuremath{\Xi^{\pm}}\xspace}
\newcommand{\Oms}          {\ensuremath{\Omega^{\pm}}\xspace}
\newcommand{\degree}       {\ensuremath{^{\rm o}}\xspace}

\newcommand{\gab}          {\ensuremath{\langle \cos(\varphi_\alpha + \varphi_\beta - 2\psid) \rangle}\xspace}

\newcommand{\dab}          {\ensuremath{\langle \cos(\varphi_\alpha - \varphi_\beta) \rangle}\xspace}

\newcommand{\gc}           {\ensuremath{\gamma_{\alpha\beta}}\xspace}

\newcommand{\dc}         {\ensuremath{\delta_{\alpha\beta}}\xspace}
\newcommand{\psid}         {\ensuremath{\Psi_{\rm 2}}\xspace}

\newcommand{\fcme}         {\ensuremath{f_{\rm CME}}\xspace}
\newcommand{\fcmexe}         {\ensuremath{f_{\rm CME}^{\rm Xe-Xe}}\xspace}
\newcommand{\fcmepb}         {\ensuremath{f_{\rm CME}^{\rm Pb-Pb}}\xspace}

\newcommand{\todo}[1]{\textcolor{darkgreen}{[TODO: \emph{#1}]}}
\newcommand{\done}[1]{\textcolor{blue}{[DONE: \emph{#1}]}}
\newcommand{\new}[1]{\textcolor{red}{#1}}
\newcommand{\old}[1]{\textcolor{orange}{#1}}
\newcommand{\change}[1]{\textcolor{blue}{[CHANGE: #1]}}
\newcommand{\ask}[1]{\textcolor{magenta}{\bf [ #1]}}
\definecolor{darkgreen}{rgb}{0.0, 0.6, 0.22}
\definecolor{dblue}{rgb}{0.,0.,1.}       
\definecolor{orange}{cmyk}{0.,0.453,1.,0.}    
\newcommand{\blue}[1]{\textcolor[named]{Blue}{#1}} 
\newcommand{\red}[1]{\textcolor[named]{red}{#1}}
\newcommand{\magenta}[1]{\textcolor[named]{magenta}{#1}}
\newcommand{\green}[1]{\textcolor[named]{Green}{#1}}
\newcommand{\orange}[1]{\textcolor{orange}{#1}}
\newcommand{\dblue}[1]{\textcolor{dblue}{#1}}
\newcommand{\dgreen}[1]{\textcolor{dgreen}{#1}}
\def \newsv {\dblue}

%% file: fa_2022-10-11_Opt_C.tex

The ALICE Collaboration would like to thank all its engineers and technicians for their invaluable contributions to the construction of the experiment and the CERN accelerator teams for the outstanding performance of the LHC complex.
The ALICE Collaboration gratefully acknowledges the resources and support provided by all Grid centres and the Worldwide LHC Computing Grid (WLCG) collaboration.
The ALICE Collaboration acknowledges the following funding agencies for their support in building and running the ALICE detector:
A. I. Alikhanyan National Science Laboratory (Yerevan Physics Institute) Foundation (ANSL), State Committee of Science and World Federation of Scientists (WFS), Armenia;
Austrian Academy of Sciences, Austrian Science Fund (FWF): [M 2467-N36] and Nationalstiftung f\"{u}r Forschung, Technologie und Entwicklung, Austria;
Ministry of Communications and High Technologies, National Nuclear Research Center, Azerbaijan;
Conselho Nacional de Desenvolvimento Cient\'{\i}fico e Tecnol\'{o}gico (CNPq), Financiadora de Estudos e Projetos (Finep), Funda\c{c}\~{a}o de Amparo \`{a} Pesquisa do Estado de S\~{a}o Paulo (FAPESP) and Universidade Federal do Rio Grande do Sul (UFRGS), Brazil;
Bulgarian Ministry of Education and Science, within the National Roadmap for Research Infrastructures 2020-2027 (object CERN), Bulgaria;
Ministry of Education of China (MOEC) , Ministry of Science \& Technology of China (MSTC) and National Natural Science Foundation of China (NSFC), China;
Ministry of Science and Education and Croatian Science Foundation, Croatia;
Centro de Aplicaciones Tecnol\'{o}gicas y Desarrollo Nuclear (CEADEN), Cubaenerg\'{\i}a, Cuba;
Ministry of Education, Youth and Sports of the Czech Republic, Czech Republic;
The Danish Council for Independent Research | Natural Sciences, the VILLUM FONDEN and Danish National Research Foundation (DNRF), Denmark;
Helsinki Institute of Physics (HIP), Finland;
Commissariat \`{a} l'Energie Atomique (CEA) and Institut National de Physique Nucl\'{e}aire et de Physique des Particules (IN2P3) and Centre National de la Recherche Scientifique (CNRS), France;
Bundesministerium f\"{u}r Bildung und Forschung (BMBF) and GSI Helmholtzzentrum f\"{u}r Schwerionenforschung GmbH, Germany;
General Secretariat for Research and Technology, Ministry of Education, Research and Religions, Greece;
National Research, Development and Innovation Office, Hungary;
Department of Atomic Energy Government of India (DAE), Department of Science and Technology, Government of India (DST), University Grants Commission, Government of India (UGC) and Council of Scientific and Industrial Research (CSIR), India;
National Research and Innovation Agency - BRIN, Indonesia;
Istituto Nazionale di Fisica Nucleare (INFN), Italy;
Japanese Ministry of Education, Culture, Sports, Science and Technology (MEXT) and Japan Society for the Promotion of Science (JSPS) KAKENHI, Japan;
Consejo Nacional de Ciencia (CONACYT) y Tecnolog\'{i}a, through Fondo de Cooperaci\'{o}n Internacional en Ciencia y Tecnolog\'{i}a (FONCICYT) and Direcci\'{o}n General de Asuntos del Personal Academico (DGAPA), Mexico;
Nederlandse Organisatie voor Wetenschappelijk Onderzoek (NWO), Netherlands;
The Research Council of Norway, Norway;
Commission on Science and Technology for Sustainable Development in the South (COMSATS), Pakistan;
Pontificia Universidad Cat\'{o}lica del Per\'{u}, Peru;
Ministry of Education and Science, National Science Centre and WUT ID-UB, Poland;
Korea Institute of Science and Technology Information and National Research Foundation of Korea (NRF), Republic of Korea;
Ministry of Education and Scientific Research, Institute of Atomic Physics, Ministry of Research and Innovation and Institute of Atomic Physics and Universitatea Nationala de Stiinta si Tehnologie Politehnica Bucuresti, Romania;
Ministry of Education, Science, Research and Sport of the Slovak Republic, Slovakia;
National Research Foundation of South Africa, South Africa;
Swedish Research Council (VR) and Knut \& Alice Wallenberg Foundation (KAW), Sweden;
European Organization for Nuclear Research, Switzerland;
Suranaree University of Technology (SUT), National Science and Technology Development Agency (NSTDA) and National Science, Research and Innovation Fund (NSRF via PMU-B B05F650021), Thailand;
Turkish Energy, Nuclear and Mineral Research Agency (TENMAK), Turkey;
National Academy of  Sciences of Ukraine, Ukraine;
Science and Technology Facilities Council (STFC), United Kingdom;
National Science Foundation of the United States of America (NSF) and United States Department of Energy, Office of Nuclear Physics (DOE NP), United States of America.
In addition, individual groups or members have received support from:
Marie Sk\l{}odowska Curie, European Research Council, Strong 2020 - Horizon 2020 (grant nos. 950692, 824093, 896850), European Union;
ICSC - Centro Nazionale di Ricerca in High Performance Computing, Big Data and Quantum Computing, European Union - NextGenerationEU;
Academy of Finland (Center of Excellence in Quark Matter) (grant nos. 346327, 346328), Finland;
Programa de Apoyos para la Superaci\'{o}n del Personal Acad\'{e}mico, UNAM, Mexico.

%% file: 2022-10-11-Alice_Authorlist_2022-10-11_Opt_C.tex
\begin{flushleft} 
\small

S.~Acharya\,\orcidlink{0000-0002-9213-5329}\,$^{\rm 126}$, 
D.~Adamov\'{a}\,\orcidlink{0000-0002-0504-7428}\,$^{\rm 86}$, 
A.~Adler$^{\rm 70}$, 
G.~Aglieri Rinella\,\orcidlink{0000-0002-9611-3696}\,$^{\rm 32}$, 
M.~Agnello\,\orcidlink{0000-0002-0760-5075}\,$^{\rm 29}$, 
N.~Agrawal\,\orcidlink{0000-0003-0348-9836}\,$^{\rm 51}$, 
Z.~Ahammed\,\orcidlink{0000-0001-5241-7412}\,$^{\rm 134}$, 
S.~Ahmad\,\orcidlink{0000-0003-0497-5705}\,$^{\rm 15}$, 
S.U.~Ahn\,\orcidlink{0000-0001-8847-489X}\,$^{\rm 71}$, 
I.~Ahuja\,\orcidlink{0000-0002-4417-1392}\,$^{\rm 37}$
% ...

\section*{Affiliation Notes}

$^{\rm I}$ Deceased\\
$^{\rm II}$ Also at: Max-Planck-Institut fur Physik, Munich, Germany\\
$^{\rm III}$ Also at: Italian National Agency for New Technologies, Energy and Sustainable Economic Development (ENEA), Bologna, Italy\\
$^{\rm IV}$ Also at: Dipartimento DET del Politecnico di Torino, Turin, Italy\\
$^{\rm V}$ Also at: Yildiz Technical University, Istanbul, T\"{u}rkiye\\
$^{\rm VI}$ Also at: Department of Applied Physics, Aligarh Muslim University, Aligarh, India\\
$^{\rm VII}$ Also at: Institute of Theoretical Physics, University of Wroclaw, Poland\\
$^{\rm VIII}$ Also at: An institution covered by a cooperation agreement with CERN\\

\section*{Collaboration Institutes}

$^{1}$ A.I. Alikhanyan National Science Laboratory (Yerevan Physics Institute) Foundation, Yerevan, Armenia\\
$^{2}$ AGH University of Krakow, Cracow, Poland\\
$^{3}$ Bogolyubov Institute for Theoretical Physics, National Academy of Sciences of Ukraine, Kiev, Ukraine\\
$^{4}$ Bose Institute, Department of Physics  and Centre for Astroparticle Physics and Space Science (CAPSS), Kolkata, India\\
$^{5}$ California Polytechnic State University, San Luis Obispo, California, United States\\
$^{6}$ Central China Normal University, Wuhan, China\\
$^{7}$ Centro de Aplicaciones Tecnol\'{o}gicas y Desarrollo Nuclear (CEADEN), Havana, Cuba\\
$^{8}$ Centro de Investigaci\'{o}n y de Estudios Avanzados (CINVESTAV), Mexico City and M\'{e}rida, Mexico\\
$^{9}$ Chicago State University, Chicago, Illinois, United States\\
$^{10}$ China Institute of Atomic Energy, Beijing, China\\
$^{11}$ Chungbuk National University, Cheongju, Republic of Korea\\
$^{12}$ Comenius University Bratislava, Faculty of Mathematics, Physics and Informatics, Bratislava, Slovak Republic\\
$^{13}$ COMSATS University Islamabad, Islamabad, Pakistan\\
$^{14}$ Creighton University, Omaha, Nebraska, United States\\
$^{15}$ Department of Physics, Aligarh Muslim University, Aligarh, India\\
$^{16}$ Department of Physics, Pusan National University, Pusan, Republic of Korea\\
$^{17}$ Department of Physics, Sejong University, Seoul, Republic of Korea\\
$^{18}$ Department of Physics, University of California, Berkeley, California, United States\\
$^{19}$ Department of Physics, University of Oslo, Oslo, Norway\\
$^{20}$ Department of Physics and Technology, University of Bergen, Bergen, Norway\\
$^{21}$ Dipartimento di Fisica, Universit\`{a} di Pavia, Pavia, Italy\\
$^{22}$ Dipartimento di Fisica dell'Universit\`{a} and Sezione INFN, Cagliari, Italy\\
$^{23}$ Dipartimento di Fisica dell'Universit\`{a} and Sezione INFN, Trieste, Italy\\
$^{24}$ Dipartimento di Fisica dell'Universit\`{a} and Sezione INFN, Turin, Italy\\
$^{25}$ Dipartimento di Fisica e Astronomia dell'Universit\`{a} and Sezione INFN, Bologna, Italy\\
$^{26}$ Dipartimento di Fisica e Astronomia dell'Universit\`{a} and Sezione INFN, Catania, Italy\\
$^{27}$ Dipartimento di Fisica e Astronomia dell'Universit\`{a} and Sezione INFN, Padova, Italy\\
$^{28}$ Dipartimento di Fisica `E.R.~Caianiello' dell'Universit\`{a} and Gruppo Collegato INFN, Salerno, Italy\\
$^{29}$ Dipartimento DISAT del Politecnico and Sezione INFN, Turin, Italy\\
$^{30}$ Dipartimento di Scienze MIFT, Universit\`{a} di Messina, Messina, Italy\\
$^{31}$ Dipartimento Interateneo di Fisica `M.~Merlin' and Sezione INFN, Bari, Italy\\
$^{32}$ European Organization for Nuclear Research (CERN), Geneva, Switzerland\\
$^{33}$ Faculty of Electrical Engineering, Mechanical Engineering and Naval Architecture, University of Split, Split, Croatia\\
$^{34}$ Faculty of Engineering and Science, Western Norway University of Applied Sciences, Bergen, Norway\\
$^{35}$ Faculty of Nuclear Sciences and Physical Engineering, Czech Technical University in Prague, Prague, Czech Republic\\
$^{36}$ Faculty of Physics, Sofia University, Sofia, Bulgaria\\
$^{37}$ Faculty of Science, P.J.~\v{S}af\'{a}rik University, Ko\v{s}ice, Slovak Republic\\
$^{38}$ Frankfurt Institute for Advanced Studies, Johann Wolfgang Goethe-Universit\"{a}t Frankfurt, Frankfurt, Germany\\
$^{39}$ Fudan University, Shanghai, China\\
$^{40}$ Gangneung-Wonju National University, Gangneung, Republic of Korea\\
$^{41}$ Gauhati University, Department of Physics, Guwahati, India\\
$^{42}$ Helmholtz-Institut f\"{u}r Strahlen- und Kernphysik, Rheinische Friedrich-Wilhelms-Universit\"{a}t Bonn, Bonn, Germany\\
$^{43}$ Helsinki Institute of Physics (HIP), Helsinki, Finland\\
$^{44}$ High Energy Physics Group,  Universidad Aut\'{o}noma de Puebla, Puebla, Mexico\\
$^{45}$ Horia Hulubei National Institute of Physics and Nuclear Engineering, Bucharest, Romania\\
$^{46}$ HUN-REN Wigner Research Centre for Physics, Budapest, Hungary\\
$^{47}$ Indian Institute of Technology Bombay (IIT), Mumbai, India\\
$^{48}$ Indian Institute of Technology Indore, Indore, India\\
$^{49}$ INFN, Laboratori Nazionali di Frascati, Frascati, Italy\\
$^{50}$ INFN, Sezione di Bari, Bari, Italy\\
$^{51}$ INFN, Sezione di Bologna, Bologna, Italy\\
$^{52}$ INFN, Sezione di Cagliari, Cagliari, Italy\\
$^{53}$ INFN, Sezione di Catania, Catania, Italy\\
$^{54}$ INFN, Sezione di Padova, Padova, Italy\\
$^{55}$ INFN, Sezione di Pavia, Pavia, Italy\\
$^{56}$ INFN, Sezione di Torino, Turin, Italy\\
$^{57}$ INFN, Sezione di Trieste, Trieste, Italy\\
$^{58}$ Inha University, Incheon, Republic of Korea\\
$^{59}$ Institute for Gravitational and Subatomic Physics (GRASP), Utrecht University/Nikhef, Utrecht, Netherlands\\
$^{60}$ Institute of Experimental Physics, Slovak Academy of Sciences, Ko\v{s}ice, Slovak Republic\\
$^{61}$ Institute of Physics, Homi Bhabha National Institute, Bhubaneswar, India\\
$^{62}$ Institute of Physics of the Czech Academy of Sciences, Prague, Czech Republic\\
$^{63}$ Institute of Space Science (ISS), Bucharest, Romania\\
$^{64}$ Institut f\"{u}r Kernphysik, Johann Wolfgang Goethe-Universit\"{a}t Frankfurt, Frankfurt, Germany\\
$^{65}$ Instituto de Ciencias Nucleares, Universidad Nacional Aut\'{o}noma de M\'{e}xico, Mexico City, Mexico\\
$^{66}$ Instituto de F\'{i}sica, Universidade Federal do Rio Grande do Sul (UFRGS), Porto Alegre, Brazil\\
$^{67}$ Instituto de F\'{\i}sica, Universidad Nacional Aut\'{o}noma de M\'{e}xico, Mexico City, Mexico\\
$^{68}$ iThemba LABS, National Research Foundation, Somerset West, South Africa\\
$^{69}$ Jeonbuk National University, Jeonju, Republic of Korea\\
$^{70}$ Johann-Wolfgang-Goethe Universit\"{a}t Frankfurt Institut f\"{u}r Informatik, Fachbereich Informatik und Mathematik, Frankfurt, Germany\\
$^{71}$ Korea Institute of Science and Technology Information, Daejeon, Republic of Korea\\
$^{72}$ KTO Karatay University, Konya, Turkey\\
$^{73}$ Laboratoire de Physique Subatomique et de Cosmologie, Universit\'{e} Grenoble-Alpes, CNRS-IN2P3, Grenoble, France\\
$^{74}$ Lawrence Berkeley National Laboratory, Berkeley, California, United States\\
$^{75}$ Lund University Department of Physics, Division of Particle Physics, Lund, Sweden\\
$^{76}$ Nagasaki Institute of Applied Science, Nagasaki, Japan\\
$^{77}$ Nara Women{'}s University (NWU), Nara, Japan\\
$^{78}$ National and Kapodistrian University of Athens, School of Science, Department of Physics , Athens, Greece\\
$^{79}$ National Centre for Nuclear Research, Warsaw, Poland\\
$^{80}$ National Institute of Science Education and Research, Homi Bhabha National Institute, Jatni, India\\
$^{81}$ National Nuclear Research Center, Baku, Azerbaijan\\
$^{82}$ National Research and Innovation Agency - BRIN, Jakarta, Indonesia\\
$^{83}$ Niels Bohr Institute, University of Copenhagen, Copenhagen, Denmark\\
$^{84}$ Nikhef, National institute for subatomic physics, Amsterdam, Netherlands\\
$^{85}$ Nuclear Physics Group, STFC Daresbury Laboratory, Daresbury, United Kingdom\\
$^{86}$ Nuclear Physics Institute of the Czech Academy of Sciences, Husinec-\v{R}e\v{z}, Czech Republic\\
$^{87}$ Oak Ridge National Laboratory, Oak Ridge, Tennessee, United States\\
$^{88}$ Ohio State University, Columbus, Ohio, United States\\
$^{89}$ Physics department, Faculty of science, University of Zagreb, Zagreb, Croatia\\
$^{90}$ Physics Department, Panjab University, Chandigarh, India\\
$^{91}$ Physics Department, University of Jammu, Jammu, India\\
$^{92}$ Physics Program and International Institute for Sustainability with Knotted Chiral Meta Matter (SKCM2), Hiroshima University, Hiroshima, Japan\\
$^{93}$ Physikalisches Institut, Eberhard-Karls-Universit\"{a}t T\"{u}bingen, T\"{u}bingen, Germany\\
$^{94}$ Physikalisches Institut, Ruprecht-Karls-Universit\"{a}t Heidelberg, Heidelberg, Germany\\
$^{95}$ Physik Department, Technische Universit\"{a}t M\"{u}nchen, Munich, Germany\\
$^{96}$ Politecnico di Bari and Sezione INFN, Bari, Italy\\
$^{97}$ Research Division and ExtreMe Matter Institute EMMI, GSI Helmholtzzentrum f\"ur Schwerionenforschung GmbH, Darmstadt, Germany\\
$^{98}$ Saga University, Saga, Japan\\
$^{99}$ Saha Institute of Nuclear Physics, Homi Bhabha National Institute, Kolkata, India\\
$^{100}$ School of Physics and Astronomy, University of Birmingham, Birmingham, United Kingdom\\
$^{101}$ Secci\'{o}n F\'{\i}sica, Departamento de Ciencias, Pontificia Universidad Cat\'{o}lica del Per\'{u}, Lima, Peru\\
$^{102}$ Stefan Meyer Institut f\"{u}r Subatomare Physik (SMI), Vienna, Austria\\
$^{103}$ SUBATECH, IMT Atlantique, Nantes Universit\'{e}, CNRS-IN2P3, Nantes, France\\
$^{104}$ Sungkyunkwan University, Suwon City, Republic of Korea\\
$^{105}$ Suranaree University of Technology, Nakhon Ratchasima, Thailand\\
$^{106}$ Technical University of Ko\v{s}ice, Ko\v{s}ice, Slovak Republic\\
$^{107}$ The Henryk Niewodniczanski Institute of Nuclear Physics, Polish Academy of Sciences, Cracow, Poland\\
$^{108}$ The University of Texas at Austin, Austin, Texas, United States\\
$^{109}$ Universidad Aut\'{o}noma de Sinaloa, Culiac\'{a}n, Mexico\\
$^{110}$ Universidade de S\~{a}o Paulo (USP), S\~{a}o Paulo, Brazil\\
$^{111}$ Universidade Estadual de Campinas (UNICAMP), Campinas, Brazil\\
$^{112}$ Universidade Federal do ABC, Santo Andre, Brazil\\
$^{113}$ Universitatea Nationala de Stiinta si Tehnologie Politehnica Bucuresti, Bucharest, Romania\\
$^{114}$ University of Cape Town, Cape Town, South Africa\\
$^{115}$ University of Houston, Houston, Texas, United States\\
$^{116}$ University of Jyv\"{a}skyl\"{a}, Jyv\"{a}skyl\"{a}, Finland\\
$^{117}$ University of Kansas, Lawrence, Kansas, United States\\
$^{118}$ University of Liverpool, Liverpool, United Kingdom\\
$^{119}$ University of Science and Technology of China, Hefei, China\\
$^{120}$ University of South-Eastern Norway, Kongsberg, Norway\\
$^{121}$ University of Tennessee, Knoxville, Tennessee, United States\\
$^{122}$ University of the Witwatersrand, Johannesburg, South Africa\\
$^{123}$ University of Tokyo, Tokyo, Japan\\
$^{124}$ University of Tsukuba, Tsukuba, Japan\\
$^{125}$ Universit\"{a}t M\"{u}nster, Institut f\"{u}r Kernphysik, M\"{u}nster, Germany\\
$^{126}$ Universit\'{e} Clermont Auvergne, CNRS/IN2P3, LPC, Clermont-Ferrand, France\\
$^{127}$ Universit\'{e} de Lyon, CNRS/IN2P3, Institut de Physique des 2 Infinis de Lyon, Lyon, France\\
$^{128}$ Universit\'{e} de Strasbourg, CNRS, IPHC UMR 7178, F-67000 Strasbourg, France, Strasbourg, France\\
$^{129}$ Universit\'{e} Paris-Saclay, Centre d'Etudes de Saclay (CEA), IRFU, D\'{e}partment de Physique Nucl\'{e}aire (DPhN), Saclay, France\\
$^{130}$ Universit\'{e}  Paris-Saclay, CNRS/IN2P3, IJCLab, Orsay, France\\
$^{131}$ Universit\`{a} degli Studi di Foggia, Foggia, Italy\\
$^{132}$ Universit\`{a} del Piemonte Orientale, Vercelli, Italy\\
$^{133}$ Universit\`{a} di Brescia, Brescia, Italy\\
$^{134}$ Variable Energy Cyclotron Centre, Homi Bhabha National Institute, Kolkata, India\\
$^{135}$ Warsaw University of Technology, Warsaw, Poland\\
$^{136}$ Wayne State University, Detroit, Michigan, United States\\
$^{137}$ Yale University, New Haven, Connecticut, United States\\
$^{138}$ Yonsei University, Seoul, Republic of Korea\\
$^{139}$  Zentrum  f\"{u}r Technologie und Transfer (ZTT), Worms, Germany\\
$^{140}$ Affiliated with an institute covered by a cooperation agreement with CERN\\
$^{141}$ Affiliated with an international laboratory covered by a cooperation agreement with CERN.\\

\end{flushleft} 